# Physics of Flow Instability and Turbulent Transition in Shear Flows

Hua-Shu Dou

Temasek Laboratories, National University of Singapore, Singapore 117411
Currently, Faculty of Mechanical Engineering and Automation, Zhejiang Sci-Tech University,
Hangzhou, Zhejiang 310018, China
Email: huashudou@yahoo.com

**Abstract**: Turbulent transition is of great significance in modern sciences and industrial applications. The physics of flow instability and turbulent transition in shear flows is studied by analyzing the energy variation of fluid particles under the interaction of base flow with a disturbance. A simple model derived from physics is proposed to show that the flow instability under finite amplitude disturbance leads to turbulent transition. It is demonstrated that it is the transverse energy gradient that leads to the disturbance amplification while the disturbance is damped by the energy loss due to viscosity along the streamline. The threshold of disturbance amplitude obtained is scaled with the Reynolds number by an exponent of -1, which is in good agreement with experiments in literature for pipe flow with injection disturbance. Experimental data for wall bounded parallel flows indicate that the critical value of the so called energy gradient parameter $K_{max}$ is *same* at turbulent transition (at least for pressure driven flows). The location of instability initiation accords well with the experiments for both pipe Poiseuille flow (r/R=0.58) and plane Poiseuille flow (y/h=0.58). It is also inferred from the proposed method that the transverse energy gradient can serve as the power for the self-sustaining process of wall bounded turbulence. Finally, the relation of "energy gradient method" to the classical "energy method" based on Rayleigh-Orr equation is also discussed.

**Keywords**: Turbulent transition; Shear flows; Energy gradient; Energy loss; Critical Reynolds number.

PACS numbers: 47.20.Ft; 47.20.Gv; 47.27.Cn; 47.15.Fe

## INTRODUCTION

Turbulence is one of the most difficult problems in classical physics and mechanics. Turbulence research has a history of more than 120 years, since Reynolds' pioneer work on the pipe flow was done (Reynolds, 1883). Reynolds showed via experiments that a nominally laminar pipe flow would display turbulent behaviour when the Reynolds number exceeded a critical value. The physical mechanisms that cause laminar flow to lose its stability and to transit to turbulence are still poorly understood (Trefethen et al, 1993; Grossmann, 2000; Drazin and Reid, 2004). From mathematical analysis, Lin (1955) demostrated that transition from laminar to turbulent flows may be due to the occurrence of instability. Emmons (1951) found the turbulent spot for the first time in experiment for natural transition of a boundary layer. His measurement indicated that the turbulent spot is the initial stage of turbulent transition and it is specifically a local phenomenon. There is intermittence at the edge of the spot surround by laminar flow. Theodorsen (1952) proposed a simple vortex model as the central element of the turbulence generation in shear flows. It takes the form of a hairpin (or horseshoe)-shaped vertical structure inclined in the direction of mean shear. Kline et al (1967) found the detailed coherent structure in the flow of a boundary layer that turbulence consists of a series of hairpin vortices. These phenomena have been confirmed by later simulations and experiments (Perry and Chong,1982; Robinson, 1991; Adrian et al, 2000). However, the challenge remains to identify the mechanisms of the formation of velocity inflection and the lift and breakdown, of hairpin vortices.

On the other hand, for Poiseuille flow in a straight pipe and plane Couette flow, linear stability analysis shows that they are stable for all the range of Reynolds number while they both transit to turbulence at finite Reynolds number in experiments (Trefethen et al, 1993; Grossmann, 2000; Drazin and Reid, 2004). Now, it is generally accepted from experiments that there is a critical Reynolds number $\text{Re}_c$ below which no turbulence can be produced regardless of the level of imposed disturbance. From experiments the critical value of the Reynolds number ($\text{Re}_c$) for pipe Poiseuille flow is approximately 2000 (Patel and Head, 1969). Above this critical value, the transition to turbulence depends to a large extent on the initial disturbance to the flow. For example, experiments showed that if the disturbance in a laminar flow can be carefully reduced, the onset of turbulence can be delayed to Reynolds number up to $\text{Re} = O(10^5)$ (Pfenninger,1961; Nishioka et al, 1975). Experiments also showed that for $\text{Re} > \text{Re}_c$, only when a threshold of disturbance amplitude is reached, can the flow transition to turbulence occur (Darbyshaire and Mullin, 1995). Trefethen et al. suggested that the critical amplitude of the disturbance leading to transition varies broadly with the Reynolds number and is associated with an exponent rule of the form, $A \propto \text{Re}^\gamma$ (Trefethen et al, 1993). The magnitude of this exponent has significant implication for turbulence research (Grossmann, 2000). In Waleffe (1995), it is shown that an exponent strictly less than -1 would indicate the importance of transient growth while -1 is expected from a simple balance between nonlinear inertial term and viscous dissipation term. Chapman, through a formal asymptotic analysis of the Navier-Stokes equations (for $\text{Re} \to \infty$), found $\gamma = -3/2$ and -5/4 for plane Poiseuille flow with streamwise mode and oblique mode, respectively, with generating a secondary instability, and $\gamma = -1$ for plane Couette flow with above both modes. He also examined the boot-strapping route to transition without needing to generate a secondary instability, and found $\gamma = -1$ for both plane Poiseuille flow and plane Couette flow (Chapman, 2002). Recently, Hof et al. (2003), used pulsed disturbances in experiments, to have obtained the normalized disturbance flow rate in the pipe for the turbulent transition, and found it to be inversely proportional to the Re number, i.e., $\gamma = -1$. This experimental result means that the product of the amplitude of the disturbance and the Reynolds number is a constant for the transition to turbulence. This phenomenon must have its physical background, and the physical mechanism of this result has not been explained so far. This issue will be clarified in the present work.

More recently, Dou (2006) suggested a new approach to analyze flow instability and turbulence transition based on the *energy gradient* concept. He proposed a function of energy gradient and then took the maximum of this function in the flow field, $K_{max}$, as the criterion for flow instability. This approach obtains a consistent value of $K_{max}$ for the critical condition (i.e., minimum Reynolds number) of turbulent transition in plane Poiseuille flow, pipe Poiseuille flow and plane Couette flow (Dou, 2006; Dou and Khoo, 2011). For flows of $K_{max}$ below this value no turbulence can be generated no matter how large of the disturbance amplitude. However, in the previous work, the detail of amplification of the disturbance by the energy gradient has not been described, and why the proposed parameter, $K_{max}$, should be used to characterize the critical condition of turbulent transition has not been derived rigorously.

In this paper, based on the analysis of disturbance of the fluid particle in shear flows, a model with support of detailed physical background and detailed derivation for flow instability and turbulent transition is proposed. In this model, the basic principle of flow instability under a disturbance has been described within the frame of Newtonian mechanics. We name the proposed model "Energy Gradient Method." With this method, the mechanism of amplification or decay of a disturbance in shear flows is elucidated. A formulation for the scaling of normalized amplitude of disturbance is obtained. Following that, the model is compared to the experimental results of others in literature.

**ENERGY GRADIENT METHOD**

In this section, we will use the basic principles of physics and mechanics to analyze the energy variation of a perturbed particle and to obtain the criterion for flow instability.

In mechanics, *instability* means that a system may leave its original rest state when disturbed. *Transition* means that a flow state has changed from laminar to turbulent, or has transited to another laminar flow state. It is not yet clear exactly how transition is related to stability (White, 1991). Linear stability theory only describes the stability of a system that has undergone an infinitesimal disturbance. In nature, a mechanical system may be stable to infinitesimal disturbance (linear stability), but can still be unstable when a sufficiently large disturbance with finite amplitude (nonlinear unstable), as shown in Fig.1. Three simple cases are demonstrated in Fig.1a to Fig.1c; a smooth ball lies at rest under stable (1a), unstable (1b) and neutral stable conditions (1c). A more complicated case is illustrated in Fig.1d, where the ball is stable for small displacement but will diverge if the disturbance is larger than a finite threshold. In fluid flow, the situation of stability is more complicated. But, we can understand it better by referring to these simple cases of stability problems in principle.

From the classical theory of Brownian motion, the microscopic particles suspended in a fluid are in a state of thermally driven, random motion (Einstein, 1956). The fluid particles exchange momentum and energy via collisions. Fig. 2 demonstrates the transport of momentum through the *layers* in parallel shear flow. Particles on neighboring layers collide, resulting in an exchange of momentum. The viscous nature of the fluids considered here, results in inelastic collision and an associated dissipation of energy. In the flow, this energy loss due to viscosity leads to the drop of total energy along the streamline. Meanwhile, there is an energy gradient in transverse direction in shear flows. The variations of energy in transverse and streamwise directions might make the disturbed particle leave its equilibrium position and forms the source (genesis) of flow instability.

Firstly, let us consider a fluid particle in the middle layer (Fig.2). This particle acquires energy from the upper layer through momentum exchange (inelastic collisions) which is expressed as $\Delta E_1$. Simultaneously, this particle releases energy to the lower layer through momentum exchange which is expressed as $\Delta E_2$. The net energy obtained by this particle from the upper and lower layer is $\Delta E = \Delta E_1 - \Delta E_2 > 0$. There is energy loss due to viscosity friction on the two interfaces, and this energy loss is expressed as $\Delta H$ ($\Delta H > 0$). For steady laminar flow,

$$\Delta E - \Delta H = 0. \tag{1}$$

Thus, the flow of this particle is in an equilibrium state. If the particle is subjected to a vertical disturbance, we then have,

$$\Delta E - \Delta H \neq 0, \tag{2}$$

and there is possibility of instability. If the particle can return its original streamline, it is in a stable equilibrium, and if it cannot, the particle is in an unstable equilibrium. For a minute displacement of the particle to the upper layer, there is $\Delta E - \Delta H > 0$, or $\Delta E / \Delta H > 1$. For a minute displacement of the particle to the lower layer, there is $\Delta E - \Delta H < 0$, or $\Delta E / \Delta H < 1$. In Fig.2, we express that the kinetic energy of this particle in steady flow is $(1/2)mu^2$ and the kinetic energy after the displacement is $\frac{1}{2}mu'^2$ where $m$ is the mass of the particle, and $u$ and $u'$ represent the velocity before and after displacement, respectively. For steady laminar flow, $\Delta E - \Delta H = 0$ corresponds to $\frac{1}{2}mu'^2 - \frac{1}{2}mu^2 = 0$, i.e. particles remain in their respective layers. For the displacement of the particle to the upper layer, $\Delta E - \Delta H > 0$ corresponds to $\frac{1}{2}mu'^2 - \frac{1}{2}mu^2 > 0$. This means that the kinetic energy of the particle will increase after being subjected to this disturbance. For the displacement of the particle to the lower layer, $\Delta E - \Delta H < 0$ corresponds to $\frac{1}{2}mu'^2 - \frac{1}{2}mu^2 < 0$. This means that the disturbance has resulted in a loss of kinetic energy for the particle. From these discussions, it is seen that the stability of a flow depends on the relative magnitude of $\Delta E$ and $\Delta H$.

Then, secondly, let us consider the *elastic* collision of particles when a disturbance is imposed to the base of a parallel shear flow (Fig.3). Let us consider that a fluid particle $P$ at its equilibrium position will move a cycle in vertical direction under a vertical disturbance, and it will have two collisions with two particles ($P_1$ and $P_2$) at its maximum disturbance distances, respectively. The masses of the three particles are $m$, $m_1$ and $m_2$, and the corresponding velocities prior to collisions are $u$, $u_1$ and $u_2$. We use primes for the corresponding quantities after collision. Without lose of generality, we may assume $m = m_1 = m_2$ for convenience of the derivation. For a cycle of disturbances, the fluid particle may absorb energy by collision in the first half-period and it may release energy in the second half-period because of the gradient of the velocity profile. The total momentum and kinetic energy are conserved during the elastic collisions. The conservation equations for the first collision on streamline $S_1$ are

$$m_1 u_1 + mu = m_1 u'_1 + mu' = \alpha_1 (m_1 + m) u_1, \tag{3}$$

and

$$\frac{1}{2} m_1 u_1^2 + \frac{1}{2} mu^2 = \frac{1}{2} m_1 u'_1{}^2 + \frac{1}{2} mu'^2 = \beta_1 \frac{1}{2} (m_1 + m) u_1^2. \tag{4}$$

Here $\alpha_1$ and $\beta_1$ are two constants and $\alpha_1 \leq 1$ and $\beta_1 \leq 1$. It should be pointed that the values of $\alpha_1$ and $\beta_1$ are not arbitrary. The values of $\alpha_1$ and $\beta_1$ are related to the residence time of the particle at $P_1$, and they are definite. If the residence time at position $P_1$ is sufficiently long (e.g. whole half-period of disturbance), the particle $P$ would have undergone a large number of collisions with other particles on this streamline and would have the same momentum and kinetic energy as those on the line of $S_1$, and it is required that $\alpha_1 = 1$ and $\beta_1 = 1$. In this case, the energy gained by the particle $P$ in the half-period is $\frac{1}{2} mu_1^2 - \frac{1}{2} mu^2$. When the particle $P$ remains on $S_1$ for less than the necessary half-period, of the disturbance, the energy gained by the particle $P$ can be written as $\beta^*_1(\frac{1}{2}mu_1^2 - \frac{1}{2}mu^2)$, where $\beta^*_1$ is a factor of fraction of a half-period with $\beta^*_1 < 1$. Similar to $\beta_1$, the value of $\beta^*_1$ is definite, its value is exactly related to the residence time of the particle at $P_1$.

The requirements of conservation of momentum and energy should also be applied for the second collision on streamline $S_2$:

$$m_2 u_2 + mu = m_2 u'_2 + mu' = \alpha_2 (m_2 + m) u_2, \tag{5}$$

and

$$\frac{1}{2}m_2 u_2^2 + \frac{1}{2}mu^2 = \frac{1}{2}m_2 u'_2{}^2 + \frac{1}{2}mu'^2 = \beta_2 \frac{1}{2}(m_2 + m) u_2^2. \tag{6}$$

Here $\alpha_2$ and $\beta_2$ are two constants and $\alpha_2 \geq 1$ and $\beta_2 \geq 1$. Similar to the first collision, $\alpha_2$ and $\beta_2$ are related to the residence time for $P$ at $P_2$. Similarly, $\alpha_2 = 1$ and $\beta_2 = 1$, when the residence time is equal to the half-period of the disturbance. For the second collision, the energy gained by particle $P$ in a half-period is $\frac{1}{2}mu_2^2 - \frac{1}{2}mu^2$ (the value is negative). The energy gained by a particle that is resident on $S_2$ for less than the half-period of the disturbance is written as $\beta^*_2(\frac{1}{2}mu_2^2 - \frac{1}{2}mu^2)$, where $\beta^*_2$ is a factor of fraction of a half-period with $\beta^*_2 < 1$. Similar to $\beta_2$, the value of $\beta^*_2$ is definite, its value is exactly related to the residence time of the particle at $P_2$.

For the first half-period, the particle gains energy by the collision and the particle also releases energy by collision in the second half-period. For the first half-cycle of the particle movement, the energy gained per unit volume of fluid is $\beta^*_1 \rho(u_1^2 - u^2)/2$. If this particle has several collisions with other particles on the path (say $N$ collisions), the energy variation of per unit volume of fluid can be written as $\sum_{i=1}^{N} \beta_1^* \rho \left( u_{1i}^2 - u^2 \right)/2$ for a half-period. If each half-period is divided into a series of intervals ($\Delta t$) and the corresponding fluid layer is a series of thin strips $\Delta y$ thick, the fraction of residence time of the particle in a layer is $\beta_1^* = \dfrac{\Delta t}{T/2}$, where $T$ is the period. If a particle stays in one layer for a full half-period, this particle will have the energy same as those within this layer, and thus $\beta_1^* = 1$. Therefore, the energy variation of per unit volume of fluid for a half-period can be written as,

$$\Delta E = \sum_{i=1}^{N} \beta_1^* \rho (u_{1i}^2 - u^2)/2 = \sum_{i=1}^{N} \frac{\Delta t}{T/2} \frac{\rho(u_{1i}^2 - u^2)/2}{y_i} y_i = \frac{2}{T} \sum_{i=1}^{N} \left( \frac{\partial E}{\partial y} \bigg|_i \right) y_i \Delta t \qquad (7)$$

where,

$$\frac{\partial E}{\partial y}\bigg|_i = \frac{\rho(u_{1i}^2 - u^2)/2}{y_i}, \qquad (8)$$

is the energy gradient in the transverse direction, and $E = (1/2)\rho u^2$ is the energy per unit volume of fluid. When $\Delta t$ tends to infinite small, Eq.(8) becomes

$$\frac{\partial E}{\partial y} = \frac{\rho(u_{1y}^2 - u^2)/2}{y}$$

at an any position of $y$ coordinate of the disturbance, and Eq.(8) becomes

$$\Delta E = \frac{2}{T} \int_0^{T/2} \frac{\partial E}{\partial y} y dt = \frac{\partial E}{\partial y} \frac{2}{T} \int_0^{T/2} y dt . \qquad (9)$$

Here, $u_{1y}$ is the velocity of mean flow at an any position of $y$ coordinate of the disturbance, and $y$ is the distance of the fluid particle deviating from its stable equilibrium position in the laminar flow. In Equation (9), $\partial E/\partial y$ is considered to be a constant in the vicinity of the streamline $S$ and also is treated as a constant in the whole cycle. This treatment can also be obtained by expanding $\partial E / \partial y$ into a Taylor series in the neighborhood of the original location and taking its leading term as a first approximation.

Without lose of generality (this will be seen later), assuming that the disturbance variation is associated with a sinusoidal function,

$$y = A\sin(\omega t + \varphi_0), \tag{10}$$

where $A$ is the amplitude of disturbance in transverse direction, $\omega$ is the frequency of the disturbance, $t$ is the time, and $\varphi_0$ is the initial phase angle. The velocity of the disturbance in the vertical direction, is the derivative of (10) with respect to time,

$$v' = \frac{dy}{dt} = v'_m \cos(\omega t + \varphi_0). \tag{11}$$

Here, $v'_m = A\omega$ is the amplitude of disturbance velocity and the disturbance has a period of $T = 2\pi/\omega$.

Substituting Eq.(10) into Eq. (9), we obtain the energy variation of per unit volume of fluid for the first half-period,

$$\begin{aligned}\Delta E &= \frac{\partial E}{\partial y}\frac{2}{T}\int_0^{T/2} y\,dt = \frac{\partial E}{\partial y}\frac{2}{T}\int_0^{T/2} A\sin(\omega t + \varphi_0)\,dt \\ &= \frac{\partial E}{\partial y}\frac{2}{T}\frac{1}{\omega}\int_0^{\pi} A\sin(\omega t + \varphi_0)\,d\omega t = \frac{\partial E}{\partial y}\frac{2A}{\pi}\end{aligned}. \tag{12}$$

The selection of the disturbance function in Eq.(10) does not affect the result of Eq.(12), except there may be a difference of a proportional constant. In a similar way, for the second half cycle, a complete similar equation to Eq.(12) can be obtained for the released energy.

Due to the viscosity of the fluid, the particle-particle collisions are more properly characterized as being inelastic. Shear stress is generated at the interface of fluid layers via momentum exchange among fluid particles, which results in energy loss. The disturbed particle is also subjected to this energy loss in a half cycle. Thus, the kinetic energy gained by a particle in a half-period is less than that represented by Eq.(12). The magnitude of the reduced part of the gained kinetic energy is related to the shear stress as well as the energy loss (see Eq.(2)).

The stability of the particle can be related to the energy gained by the particle through vertical disturbance and the energy loss due to viscosity along streamline in a half-period. It is now left to use to calculate the energy loss due to viscosity in a half-period in the following. Assuming that the streamwise distance moved by the fluid particle in a period is far less than the length of the flow geometry, the evaluation of the energy loss is derived as follows. In the half-period, the particle moves a short distance of $l$ along a streamline, thus it has an energy loss per unit volume of fluid along the streamline, $(\partial H / \partial x)l$, where $H$ is the energy loss per unit volume of fluid due to viscosity along the streamline. The streamwise length moved by the particle in a half-period can be written as, $l = u(T/2) = u(\pi/\omega)$. Thus, we obtain

$$\Delta H = \frac{\partial H}{\partial x} l = \frac{\partial H}{\partial x} \frac{\pi}{\omega} u. \tag{13}$$

Thus far, the energy variation of per unit volume of fluid for the first half-period, $\Delta E$, and the energy loss along the streamline per unit volume of fluid for the first half-period, $\Delta H$, have been obtained as in Eq.(12) and Eq.(13). In the method, we do not trace the single particle and do not give an identity to each particle, but we analyze the macro behaviour of large quantity of fluid particles in locality. Now, we discuss how a particle loses its stability by comparing the terms: $\Delta E$ and $\Delta H$. After the particle moves a half cycle, if the net energy gained by collisions is zero, this particle will stay in its original equilibrium position (streamline). If the net energy gained by collisions is larger than zero, this particle will be able to move into equilibrium with a higher energy state. If the collision in a half-period results in a drop of kinetic energy, the particle can move into lower energy equilibrium. However, there is a critical value of energy increment which is balanced (damped) by the energy loss due to viscosity (see Eq.(2) and the discussion). When the energy increment accumulated by the particle is less than this critical value, the particle could not leave its original equilibrium position after a half-cycle. Only when the energy increment accumulated by the particle exceeds this critical value, could the particle migrate to its neighbor streamline and its equilibrium will become unstable (we can understand better with reference to Fig.1d). Therefore, if the net energy gained by collisions is less than this critical amount, the disturbance will be damped by the viscous forces of the fluid and this particle will still stay (return) to its original location (Fig.4a). If the net energy gained by collisions is larger than this critical amount, this particle will become unstable and move up to neighboring streamline with higher kinetic energy (Fig.4b). Similarly, in the second half-period, if the energy released by collision is not zero, this particle will try to move to a streamline of lower kinetic energy. If the energy released by collision is larger than the critical amount, this particle becomes unstable and moves to a equilibrium position of lesser energy (Fig.4c). If the energy increments in both of the half-periods exceed the critical value, the particle would oscillate about the original equilibrium and a disturbance wave would be generated. This describes how a particle loses its stability and how the instability occurs. A continuous cycle of particle movement will lead to the particle to gradually deviate from its original location, thus an amplification of disturbance will be generated. Since linear instability is only associated with infinitesimal disturbance amplitude, it is clear from the discussion here that *it is the nonlinearity of the disturbance with finite amplitude that acts as a source for instability occurrence.*

As discussed above, the relative magnitude of the energy gained from collision and the energy loss due to viscous friction determines the disturbance amplification or decay. Thus, for a given flow, a **stability criterion** can be written as below for the half-period, by using Eq.(12) and Eq.(13),

$$F = \frac{\Delta E}{\Delta H} = \left(\frac{\partial E}{\partial y}\frac{2A}{\pi}\right) \bigg/ \left(\frac{\partial H}{\partial x}\frac{\pi}{\omega}u\right) = \frac{2}{\pi^2}K\frac{A\omega}{u} = \frac{2}{\pi^2}K\frac{v'_m}{u} < Const, \qquad (14)$$

and

$$K = \frac{\partial E / \partial y}{\partial H / \partial x}. \qquad (15)$$

Here, $F$ is a function of coordinates which expresses the ratio of the energy gained in a half-period by the particle and the energy loss due to viscosity in the half-period. $K$ is a dimensionless field variable (function) and expresses the ratio of transversal energy gradient and the rate of the energy loss along the streamline. It should be mentioned that there is no approximation in deriving Eq.(14) except that there may exist a proportional constant due to the expression of the disturbance function introduced in Eq.(10).

It can be found from Eq.(14) that the instability of a flow depends on the values of $K$ and the amplitude of the relative disturbance velocity $v'_m/u$. The magnitude of $K$ is proportional to the global Reynolds number (to be detailed later). Thus, it can be seen from Eq.(15) that $F$ increases with the Reynolds number Re. The maximum of $F$ in the flow field will reach its critical value first with the increase of Re. The critical value of $F$ indicates the onset of instability in the flow at this location and the initiation of flow *transition* to turbulence. Therefore, at the onset of turbulence, the transition from laminar to turbulent flows is a local phenomenon. It is not surprising that a turbulence spot can be observed in earlier stage of the transition process. Experiment confirmed that the turbulent spot is actually a localized turbulence phenomenon which is resulted from the hairpin vortices (Adrian et al, 2000; Singer and Joslin, 1994; Hommema and Adrian, 2002). As observed from experiments, a small region of turbulence is generated in the flow at a relatively low Re number, while the turbulence is generated in the full domain at a high Re (Drazin and Reid, 2004; Wu et al, 2006).

From Eq.(8), we have $\frac{\partial E}{\partial y} \sim \frac{\rho U^2}{L}$; The rate of energy loss of per unit volume of fluid along the streamwise direction is $\frac{\partial H}{\partial x} \sim \frac{(\tau \cdot L^2) \cdot L / L^3}{L} = \frac{\mu U / L}{L} = \mu \frac{U}{L^2}$. Here, $\tau$ is the shear stress, $\mu$ is the dynamic viscosity, $U$ is the characteristic velocity and $L$ is the characteristic length. Thus, for a given geometry and flow condition, we obtain the following equation from Eq. (15)

$$K = \frac{\partial E/\partial y}{\partial H/\partial x} \sim \frac{\rho U^2/L}{\mu U/L^2} = \text{Re}, \tag{16}$$

where $\text{Re} = \rho U L/\mu$ is the Reynolds number. For any type of flows, it can be demonstrated that the variable $K$ is proportional to the global Reynolds number for a given geometry (Dou, 2006). Therefore, the criterion of Eq.(14) can be written as,

$$\text{Re}\frac{v'_m}{u} < Const \tag{17}$$

or

$$\left(\frac{v'_m}{u}\right)_c = \frac{C_1}{\text{Re}}, \tag{18}$$

where $C_1$ is a constant. Since the disturbance of velocity at a location in the flow field can be written as,

$$\frac{v'_m}{u} = \frac{v'_m}{U}\frac{U}{u} \sim \frac{v'_m}{U}, \tag{19}$$

Eq.(18) can be written as

$$\left(\frac{v'_m}{U}\right)_c = \frac{C_2}{\text{Re}}, \tag{20}$$

where $C_2$ is another constant. Here, $(v'_m/U)_c$ is the normalized amplitude of the velocity disturbance at critical condition. In Eq.(19), the fact is used that the velocity ratio $U/u$ is a constant at a position in the flow, for example, $\frac{U}{u} = \frac{1}{2}\left(1 - \frac{r^2}{R^2}\right)^{-1}$, for pipe flow.

If the *Re* is sufficiently small (e.g., Re<2000 for pipe Poiseuille flow), the energy gained by the disturbed particle in a half-period is similarly small. Even if the disturbance amplitude is large, the particle still cannot accumulate enough energy for the stability criterion of the particle (*F* in Eq (14)) to exceed the critical number. Therefore, it can be found from Eq.(14) that there exists a critical value of the non-dimensional field variable *K* below which the flow remains laminar always. The critical value of *K* is decided by its maximum ($K_{max}$) in the domain. Thus, we take $K_{max}$ as the energy gradient parameter. The critical value for $K_{max}$ is related to the critical Reynolds number for the onset of turbulence. For situations where $K_{max}$ is below this critical value, all the energy gained by collision is damped (by viscous friction) and the flow is stable, independent of magnitude of the normalized disturbance. For a parallel flow, the fluid particles flow along straight streamlines and the energy gradient only involves kinetic energy (there is no gradient of pressure energy or potential energy). Thus, *the critical value of $K_{max}$ should be a constant for all parallel flows*. From these discussions, the physical implication of the critical Reynolds number for turbulence transition can be further understood. The critical Reynolds number is the minimum Re, below which the disturbed particle could not accumulate sufficient energy to leave its equilibrium state because the energy loss due to viscosity is large.

For pressure driven flows, the energy loss due to viscosity along the streamline equals to the magnitude of the energy gradient along the streamline (Dou,2006). Thus, for this case, the function *K* expresses the ratio between the energy gradient in the transverse direction and that in the streamwise direction. This is why the "*Energy Gradient Method*" is named. If there is an *inflection point* on the velocity profile, the energy loss $\frac{\partial H}{\partial x}$ at this point is zero. The value of function *K* becomes infinite at this point and indicates that the flow is unstable when it is subjected to a finite disturbance (see Eq.(14)). As such, the existence of inflection point on a velocity profile is a sufficient condition for flow instability. For inviscid flows, that the value of function *K* becomes infinite at the inflection point is still true. Therefore, for inviscid flows, the existence of inflection point on a velocity profile is a sufficient condition for flow instability, but not a necessary condition.

Further explanation of the physical implication of the critical value in Kmax is given as follow, from which we can better understand the model presented here. For pressure driving flow (without work input or output), we obtain, $dE = -dH$, from the Navier-Stokes equation along the streamline (Dou et al, 2007). Then, we have

$$\frac{\partial H}{\partial x} = -\frac{\partial E}{\partial x} \qquad (21)$$

Thus, for pressure drive flows, the equation (15) can be written as

$$K = \frac{\partial E/\partial y}{\partial H/\partial x} = -\frac{\partial E/\partial y}{\partial E/\partial x}. \qquad (22)$$

As we concern the magnitude of K, the minus symbol in the equation is not important.

Equation (22) indicates that the function K represents the ratio between the energy gradients in the two directions, and characterizes the direction of the vector of total energy gradient in the flow field. This is why the model is named as "Energy gradient model." The value of "$\arctan K$" expresses the angle between the direction of the total energy gradient and the streamwise direction. Therefore, we write,

$$\alpha = \arctan K. \qquad (23)$$

The angle α is named as "energy angle," as shown in Fig.5. There is a critical value of energy angle, $\alpha_c$ ($\alpha_c = \arctan K_c$), corresponding to the critical value of $K_{max}$. Thus, the value of the energy angle (its absolute value) can also be used to express the extent of the flow near the instability occurrence,

(1) $\alpha < \alpha_c$, the flow is stable.

(2) $\alpha \geq \alpha_c$, the flow is unstable.

(3) $\alpha = 90°$, the flow is unstable.

Here $\alpha_c$ is called the critical energy angle for flow transition. When $\alpha > \alpha_c$, the flow becomes unstable if it is subjected a disturbance. When $K \to \infty$, we have $\alpha \to 90°$. At this condition, the threshold of disturbance energy needed to trigger the transition at a given base flow (and thus Re is given as infinite) is infinitely small. Figure 5 show the schematic of the energy angle for plane Poiseuille flows as an example. For Poiseuille flows, $0° \leq \alpha < 90°$, the flow stability depends on the magnitude of $\alpha$ and the level of disturbance. For parallel flow with a velocity inflection, α =90° ($K_{max}=\infty$) at the inflection point, and the flow is therefore unstable (Fig.6).

**COMPARISON WITH EXPERIMENTS AND DISCUSSIONS**

**Mechanism of turbulence transition event**

It is well known that there is a coherent structure in developed turbulence (Robinson, 1991). This coherent structure consists of a series of hairpin vortices with scale of the same order as the flow geometry. This form of a hairpin (or horseshoe)-shaped vertical structure has been confirmed by extensive experiments and simulations (Adrian et al, 2000; Singer and Joslin, 1994; Hommema and Adrian, 2002). Both simulations and experiments showed that the development of the hairpin vortex in boundary layer flows will lead to the formation of the young turbulent spot (Adrian et al, 2000; Singer and Joslin, 1994; Hommema and Adrian, 2002)., which will result in the evolution of developed turbulent flow when Re is high. In the process of turbulence generation, two bursting phenomena, namely, an *ejection* and a *sweep (or in-rush)*, are generated, the former refers to the ejection of low-speed fluid from the wall, while the latter means the impinging of high-speed flow towards the wall. There must be some driving mechanisms behind these phenomena from the view point of mechanics. However, the mechanisms of these phenomena are still not well understood although there are a lot of studies for these. In present model, these phenomena in shear flows can be explained as follows: When a disturbance is imposed to the base flow, the fluid particle gains energy via the interaction of the disturbance in the transverse direction and the energy gradient of the base flow in transverse direction. If the energy variation in the cycle is much larger than the energy loss due to viscous friction and the criterion in Eq.(14) is violated, nonlinear instability will occur and the particle will move to a new equilibrium position, with an energy state that depends upon the result of the disturbance cycle on the particle. If the particle gains energy during the complete cycle of the disturbance, this particle moves to higher energy position (upward, Fig.4b), coinciding with a higher energy state. After a continuous migration upward of the fluid particle, an inflection point on the velocity profile will be produced. Further downstream, a continuous interaction of the particle and the transverse energy gradient could lead to the lift of spanwise vortex roll and the formation of hairpin vortex (this is typical for boundary layer flow, see Kline et al, 1967), which will results in a *bursting (ejection)* and the appearance of the young turbulent spot at further downstream after more disturbance amplification. As is well known, the appearance of the turbulent spot is the primary stage of the generation of turbulence (Hommema and Adrian, 2002). When the particle releases energy during the whole cycle (total energy increment is negative except viscous loss), this particle will move downward in the flow (Fig.4c), to a position with lower energy. After a few cycles of motion, this kind of events will make the flow profile swollen, compared to the normal velocity profile without disturbance. In this case, the so called *sweep (or in-rush)* will be generated after a continuous migration. Although the hairpin vortex structure was first found in boundary layer flows, hairpin-vortex packets are also found in other wall-bounded shear flows (Hommema and Adrian, 2002). From these discussions, it can be seen that how the nonlinear instability is connected to the generation of turbulence. The first occurrence of nonlinear instability actually corresponds to the beginning of the transition process, and the final transition to full turbulence is the results of a series of nonlinear interactions of the disturbance with the energy gradient of the base flow.

Jimenez and Pinelli (1999) used numerical simulations to demonstrate that a self-conservation cycle exists which is local to the near-wall region and does not depend on the outer flow for wall bounded turbulence. Waleffe (1997) also discovered a similar mechanism via detailed mathematical analysis. However, what mechanism should provide the power to drive this cycle which is independent of the outer flow? The present model explains that the transversal energy gradient plays the part of the source of the disturbance amplification and transfers the energy to the disturbance via the interaction. With a similar way as in Dou (2006), applying Eq.(15) to the Blasius boundary layer flow and calculating the distribution of $K$, it is easy to find that the position of $K_{max}$ is very near the wall (within 1/10 of the thickness of the boundary layer), see Dou and Khoo (2009). This is why the turbulence is always generated very near the wall for boundary layer flow (Kline et al, 1967, Perry and Chong, 1982). In comparison, the position of $K_{max}$ in pipe Poiseuille flow is at $r/R$=0.58 (Dou, 2006).

For the boundary layer flow, the transverse velocity is not zero (not exactly parallel flow). At a higher Re, the generation of linear instability leads to propagating of Tollmien-Schlichting waves and the formation of streamwise streaks as well as appearance of streamwise vortices (Wu et al, 2006; Drazin and Reid, 2004). These streamwise vortices make the streamwise velocity periodically "inflectional" and "swollen" along the spanwise direction (Fig.7). At such background, the nonlinear interaction of disturbed particles with transversal energy gradient will lead to the instability which results in the "ejection" at the inflection side and the "sweep" at the swollen side at larger disturbance, and further development of these events may result in transition to turbulence. In the developed turbulence, these phenomena may occur randomly in the flow. The dominating factors to lead to turbulence transition and those to sustain a turbulence flow should be the same, since there is similarity between these two types of flows found from experiments (Lee, 2000). Thus, these phenomena may also provide a mechanism for the self-sustaining process of wall bounded turbulence. The nonlinear interaction of disturbed particles with energy gradient continuously transfers the energy from the mean flow to the vortex motion in developed turbulence, therefore, turbulence is sustained.

**Threshold amplitude of disturbance scaled with Re**

Many researchers have investigated the scaling relationship between the threshold amplitude of the disturbance and the Reynolds number (Re) (Thefethen et al, 1993; Chapman, 2002). Recently, Hof et al (2004) repeated the experiment of pipe flow done 120 years ago by Reynolds (1883) with detailed care of control. It was found that the scaling is well fitted by an exponent -1, i.e., $(v'_m/U)_c$ inversely proportional with Re, as shown in Fig.8. The mechanism of this phenomenon has not been explained so far. One can find that the result in the present study (Eq.(20)), $(v'_m/U)_c \sim \text{Re}^{-1}$, obtains exactly agreement with the experiments of Hof et al (2003) and therefore the present model well explains the physics of scaling law derived from their experimental data. Shan et al. (1998)'s results of direct numerical simulation for transition in pipe flow under the influence of wall disturbances also showed that the critical amplitude of disturbance is scaled with Re by the exponent of -1. It is interesting that this exponent (-1) has also been found in experiments on transition in boundary layers (Govindarajan and Narasimha, 1991); this agreement has been discussed by Hof et al (2003).

The physical mechanism of the effect of disturbance amplitude on the stability can be concisely explained by the present model. In the first half-period of the disturbance cycle at a given Re, if the amplitude of disturbance is large, the particle could gain more energy because it can exchange energy with particles with higher kinetic energy. However, the viscous energy loss may not increase (for example, the energy loss is constant for the whole flow field in simple pipe and plane Poiseuille flows (Dou, 2006) which equals to the pressure drop per unit length). Thus, the energy gained in the cycle will be much larger than the energy loss, which leads to the flow being more unstable. Similarly, in the second half-period of the disturbance cycle, if the amplitude of disturbance is large at a given Re (say, exceeds the value expressed by Eq.(20)), the particle could release more energy because it can exchange energy with particles with lower kinetic energy. However, as stated above, the magnitude of the energy loss may not change much. Thus, the energy released can reach its threshold expressed by Eq.(20) for instability occurrence at lower Re.

**Critical value of $K_{max}$ for turbulent transition**

It is mentioned in previous section that the critical value of $K$ in Eq. (14) is decided by its maximum ($K_{max}$) in the field and should be a constant for parallel shear flows. In this section, we will give the comparison of the theory with the experimental data for the critical condition of turbulent transition for parallel flows. The derivation of function $K$ in Eq.(14) for the pipe Poiseuille flow, plane Poiseuille flow and plane Couette flow have been given previously in (Dou, 2006; Dou and Khoo, 2011). The schematic diagrams of these flows are shown in Fig.9.

For pipe Poiseuille flow, the function $K$ has been derived in (Dou, 2006), here it is just introduced below,

$$K = K\left(\frac{r}{R}\right) = \frac{1}{2}\text{Re}\frac{r}{R}\left(1 - \frac{r^2}{R^2}\right) \tag{24}$$

Here, $\text{Re} \equiv \frac{\rho U D}{\mu}$ is the Reynolds number, $\rho$ is the density, $\mu$ is the dynamic viscosity, $U$ is the averaged velocity, $r$ is in the radial direction of the cylindrical coordinate system, $R$ is the radius of the pipe, and $D$ the diameter of the pipe. It can be seen that $K$ is a cubic function of radius, and the magnitude of $K$ is proportional to Re for a fixed point in the flow field. The position of the maximum value of $K$ occurs at $r/R=0.58$.

For plane Poiseuille flow, the function *K* has been derived in Dou (2006), here it is just introduced below,

$$K = K\left(\frac{y}{h}\right) = \frac{3}{4}\text{Re}\frac{y}{h}\left(1 - \frac{y^2}{h^2}\right) \tag{25}$$

Here, $\text{Re} \equiv \frac{\rho U L}{\mu}$ is the Reynolds number, *U* is the averaged velocity, *y* is in the transversal direction of the channel, *h* is the half-width of the channel, and *L=2h* is the width of the channel. It can be seen that *K* is a cubic function of y which is similar to the case of pipe flow, and the magnitude of *K* is proportional to Re for a fixed point in the flow field. The position of the maximum value of *K* occurs at *y/h=0.58*. In references, another definition of Reynolds number is also used, $\text{Re} \equiv \frac{\rho u_0 h}{\mu}$, where $u_0$ the velocity at the mid-plane of the channel (Trefethen, 1993).

For plane Couette flow, the function *K* has been derived in Dou and Khoo (2011), here it is just introduced below,

$$K = K\left(\frac{y}{h}\right) = \text{Re}\frac{y^2}{h^2}, \tag{26}$$

where $\text{Re} \equiv \frac{\rho u_h h}{\mu}$ is the Reynolds number, $u_h$ is the velocity of the moving plate, *y* is in the transversal direction of the channel, *h* the half-width of the channel. It can be seen that *K* is a quadratic function of *y/h* across the channel width, and the magnitude of *K* is proportional to Re at any location in the flow field. The position of the maximum value of *K* occurs at *y/h=1.0*, and

$$K_{\max} = \frac{\rho U h}{\mu} = \text{Re}. \tag{27}$$

The values of $K_{max}$ at the critical condition determined by experiments for various types of flows are shown in Fig,10 and Table 1. We take this critical value of $K_{max}$ for the turbulent transition as *Kc*. It is seen that the critical value of $K_{max}$ for all the three types of flows fall within in a narrow range of 370~389. It is observed that although the critical Reynolds number is different for these flows, the critical value of $K_{max}$ is the same for these flows. This demonstrates that $K_{max}$ is really a dominating parameter for the transition to turbulence. These data strongly support the proposed method in present study and the claim that the critical value of $K_{max}$ is constant for all parallel flows, as discussed before.

For wall bounded parallel flows, Kc=370~389 corresponds to the critical energy angel $\alpha_c = 89.85°$ in Eq.(23). This means that turbulent transition can only be possible when $\alpha = 89.85° \sim 90°$. When $\alpha < 89.85°$, no turbulence exists despite of the disturbance.

In Table 1, the critical Reynolds number determined from energy method is also listed in it for purpose of comparison later. The critical Reynolds number determined from eigenvalue analysis of linearized Navier-Stokes equations is also listed for reference.

In the proposed method, the flow is expected to be more unstable in the area of high value of $K$ than that in the area of low value of $K$. In the flow field, the instability should start first at the location of maximum of $F$ according to Eq.(14) with the increase of Re. For a given disturbance, the first instability should be associated with the maximum of $K$, $K_{max}$, in the flow field if the amplitude of disturbance does not change much in the neighborhood of $K_{max}$. That is, the position of maximum of $K$ is the most unstable position. For a given flow disturbance, there is a critical value of $K_{max}$ over which the flow becomes unstable. Now, it is difficult to directly predict this critical value by model. However, it can be determined using available experimental data as done in Table 1. It is better to distinguish that $K_{max}$ is the maximum of the magnitude of $K$ in the flow domain at a given flow condition and geometry, and $K_c$ is critical value of $K_{max}$ for instability initiation for a given geometry.

Recently, Hof et al. (2004) have shown for pipe flow that there exist unstable traveling waves with computational studies of the Navier-Stokes equations and ideas from dynamical systems theory. It is suggested that traveling waves moving through the fluid at different speeds might be responsible for the onset and sustenance of turbulence. These traveling wave solutions consist of streamwise swirls and streaks with rotational symmetry about the axis of the pipe. The outlines of these solutions of traveling waves seem to obtain good agreement with experimental observations in pipe flow. Hof et al. (2004) suggested that the dynamics associated with these unstable states may indeed capture the nature of fluid turbulence. Dou suggested that these traveling waves may be associated with the instability resulting from the transverse energy gradient since the location of the kink (inflection of velocity profile) on the velocity profile obtained by the solution of traveling waves accords with the position of the maximum of the function $K$ (Dou, 2006). According to the present study, with the increase of Re, the oscillation of base flow should start first from the position of $F_{max}$, see Eq.(14). If $(v'_m/u)_{max}$ does not vary too much at the neighborhood of $F_{max}$, the position of $K_{max}$ coincides approximately with that of $F_{max}$. The oscillation of base flow could lead to secondary flows if the oscillation amplitude is large. The secondary flow should appear first around the position of $K_{max}$. The experiments by Hof et al. (2004) showed that the streamwise vortices at Re=2000 (the base flow is still laminar) occur at about $r/R$=0.5-0.6 (Fig.2(A) in Hof et al, 2004), which accords with the present study that we found the maximum of $K$ occurring at the ring of $r/R$=0.58 (Dou, 2006).

More recently, Nishi et al (2008) did experiment of turbulent transition for pipe flow through puffs and slugs generation and the disturbance was introduced at the pipe inlet by a short duration of inserted "wall fences". Figure 11 and 12 show an example of the experimental results chosen from a large number of time records for the instantaneous velocities of the puffs at different radial locations r/R. The typical axial velocity at different radial position r/R vs. time which is shown in Fig.11 from the time when the control system for disturbance operated at Re=2450. The cross-sectional velocity profiles measured at the exit of the pipe for the puff structures at Re=2450 are shown in figure 12. The structures reveal laminar-to- turbulent transition between t =4 and 4.40 s. These figures display the puff structures correspond to Re=2450. It is clearly shown that the laminar flow is basically smooth when time is from 4.0 s to 4.25 s, and the velocity oscillation first appears at r/R=0.47-0.73 from time of 4.25 s. This fact indicates that the flow becomes most unstable in the range of r/R=0.47-0.73 under the disturbance influence, which is in agreement with the prediction in this study that the r/R=0.58 is the most unstable position to first making initiation of transition.

For plane Poiseuille flow, the position of the maximum of *K* occurs at *y/h=0.58* so that this position is the most dangerous position for instability. Nishioka et al (1975)'s experimental data has shown that the flow oscillation first appears at the location of about *y/h=0.6*. For plane Couette flow, the position of the maximum of *K* occurs at *y/h=1.0.* Owing to the fact of no-slip at the wall, the disturbance at the wall is zero. The most dangerous position should be off a short distance from the wall such that the magnitude of the disturbance is apparently playing a role and the value of *K* is still large. Thus, the value of *F* could get large value and, therefore, the Eq.(14) is violated. Some nonlinear analysis showed that the development of disturbance and the distortion of base flow first start at the layer near one of the walls (Lessen and Cheifetz, 1975).

The energy gradient method has been applied to Taylor-Couette flow between concentric rotating cylinder and it is confirmed that this method is also applicable to rotating curved flows if the kinetic energy in parallel flows is replaced by the total mechanical energy (kinetic energy plus pressure energy while gravitational energy is neglected) (Dou et al , 2008). Our another previous work demonstrated that the energy loss due to viscosity along the streamline has stable role to the disturbance (Dou et al, 2007).

**Comparison of Energy gradient method with Energy method**

The critical Reynolds numbers determined from energy method are also included in Table 1 for various flows which are taken from (Drazin and Reid, 2004). It can be seen that the critical values of the Reynolds number for various flows by this method are much lower than those obtained from experiments. Energy method is based on the famous Reynolds-Orr equation (Drazin and Reid, 2004),

$$\frac{dk(t)}{dt} = -\int_V u_i u_j \frac{\partial U_i}{\partial x_j} dV - \frac{1}{\mathrm{Re}} \int_V \frac{\partial u_i}{\partial x_j} \frac{\partial u_i}{\partial x_j} dV \qquad (28)$$

where $k(t) = \frac{1}{2}\int_v u_i u_i dV$ is the kinetic energy of disturbances. This equation is integrated over the flow domain $V$. The first term on the right side of the equation is the production of disturbance kinetic energy and the second term on the right side is the dissipation of disturbance energy in the system. The term $dk/dt$ in the left side of Eq.(28) means the rate of increase of disturbance kinetic energy over the system. When $Re$ is sufficient small so that $dk/dt<0$, the flow is stable.

The energy method looks at the variation of the kinetic energy in the whole domain with the time for a given Re. The critical Reynolds number determined with it is the minimum Reynolds number below which the kinetic energy of any finite-amplitude disturbance decay monotonically (Drazin and Reid, 2004). Actually, at a given Re, the kinetic energy in the system, $k$, may first increase and then decrease with the time. When the $k$ reaches its maximum, the flow may not achieve its threshold to lose its stability. This is because the flow instability does not depend on the temporal increase of the kinetic energy of the disturbance if the disturbance amplitude is not sufficiently large (see Eq.(14)). Therefore, the critical $Re$ obtained with energy method may not be the real critical $Re$ to make the flow instability and it is generally much lower than the experimental value as shown in Table 1.

In energy gradient method, the flow instability is not based on the increase of the disturbance energy with the time. The essence of this method is to observe the stability of mean flow caused by the interaction of the disturbance with the base flow. This interaction leads to variation of the distribution of energy of mean flow. When the variation of the energy of the mean flow reaches a threshold at a position in the domain as described in previous sections, the mean flow will lose its stability due to the requirement of energy equilibrium.

However, the energy gradient method is related to the *magnitude* of the kinetic energy of the local disturbance. In Eq.(14), it is noted that the function $F$ is proportional to the disturbance amplitude, $F \propto K \frac{v'_m}{u}$, while the kinetic energy of the disturbance is generally proportional to the squares of the disturbance amplitude and the disturbance frequency, $k \propto {v'_m}^2 \omega^2$ for a normal disturbance. Thus, we have $v'_m \propto \sqrt{k/\omega^2}$ and $F \propto K \frac{\sqrt{k/\omega^2}}{u}$. Therefore, a large kinetic energy of disturbance will promote the instability when the frequency of the disturbance is fixed for a given base flow.

**Some Further remarks**

The present study shows that the flow stability is purely a stability of the mechanical energy field. The analysis of the stability is not started from the Navier-Stokes equations but based on the principle of Newtonian mechanics. Therefore, it is clear that the energy gradient model is compatible to, and is not contradicting with the Navier-Stokes equations. Actually, in order to compare the method with the experiments, the calculations of $K$ for the three types of parallel flows are obtained via the analysis of Navier-Stokes equations.

The *essence* of flow instability and turbulent transition can be understood as follow according to the proposed model. The disturbance causes variation of the energy field of mean flow and leads to the energy field of mean flow to lose its equilibrium. To reach a new equilibrium state, the flow instability must occur. The instability is the beginning of the transition, and the transition is a process of disturbance development. When this process is completed, the full developed turbulence is formed. We may restate the principle of energy gradient method in a simple way as follow. In shear flows, disturbed fluid particles wander at their original equilibrium position under the disturbance. This wandering makes their kinetic energy exchanged with others in the neighboring streamlines and leads to their kinetic energy of mean flow differing from those which are not disturbed or not largely disturbed at the original streamline. Thus, an energy difference of mean flow is formed between the disturbed particle and the undisturbed particle at the original streamline. Therefore, if this energy difference is sufficient large, these disturbed particles can migrate toward their new equilibrium position in term of energy due to the equilibrium role of energy so that these particles lose the stability. A similar and simple example is the two-phase flow in which some polymer particles are uniformly suspended in the water and flow with the water. When the temperature rises, the density of water has no change, but the density of polymer may decrease. Due to the role of equilibrium of gravitational energy, these polymer particles (like the disturbed particle in pure fluid flow) will lose their stability and will migrate up until they reach the surface of the water. The difference is that the energy leading to instability is the kinetic energy for the case of parallel flows of pure fluid, while the energy leading to instability is the gravitational energy for the case of the said two-phase flow.

The scaling of disturbance in this study is worked out under the assumption that the base flow is *parallel flow* and the fluid particle is subjected to a transversal disturbance. It is believed that for crossflow injection disturbance introduced to all parallel flows, the amplitude scales with Re by the exponent of -1, based on present analysis. The scaling is suitable for injection disturbance or natural transition for all parallel shear flows (both wall bounded flow and free shear flow). The scaling law of -1 is not suitable for push-pull disturbance in Peixinho and Mullin, 2007) where an exponent of -1.3 or -1.5 is observed and the disturbance induced by ring-type obstacles (Nishi et al, 2008). In the latter two types of disturbances induced by ring-type obstacles, the base flow is made not be a full developed laminar parabolic flow during the transition initiation (i.e., not parallel flow), while the base flow keeps to be parallel flow before breaking down for injection disturbance.

Criteria for flow instability and turbulent transition have been given under the frame of energy gradient theory (Dou, 2004; Dou, 2006; Dou, 2007; Dou and Khoo, 2010). These criteria are summarized as follow.

**Theorem (1):** Potential flow (inviscid and $\nabla \times u = 0$) is stable.

**Theorem (2):** Inviscid rotational ($\nabla \times u \neq 0$) flow is unstable.

**Theorem (3):** Velocity profile with an inflectional point is unstable when there is no work input or output to the system, for both inviscid and viscous flow, in curved streamline configurations (including parallel flow configurations).

For viscous flow, the flows can be classified as pressure driven and shear driven flows according to the energy process. That there is no work input or output to the system means the pressure driven flow.

**Theorem (4):** For pressure driven flow, the necessary condition and sufficient condition for turbulent transition is the presence of velocity inflection of the *averaged flow profile*.

**Theorem (5):** For shear driven flow, the necessary condition and sufficient condition for turbulent transition is the existence of zero velocity gradient on the velocity profile of the *averaged flow profile*.

**CONCLUSIONS**

After analyzing the process of energy transfer in perturbed shear flows we have developed a model for flow instability, called the "Energy Gradient Method". The method proposes that in shear flows it is the transverse energy gradient interacting with a disturbance to lead to the flow instability, while the energy loss, due to viscous friction along the streamline, damps the disturbance. The mechanisms of velocity inflection and formation and lift of the hairpin vortex in shear flows are well explained with the analytical result; the disturbed particle exchanges energy with other particles in base flow in transverse direction during the cycle and causes the particle leaves its equilibrium position.

The proposed theoretical model is in agreement with the experiments primarily in three aspects: (1) The threshold amplitude of disturbance for transition to turbulence is scaled with Re by an exponent of -1 in parallel flows, which explains the recent experimental results of pipe flow by Hof et al. (2003) and also Peixinho and Mullin (2007) where injection disturbances are used. (2) For wall bounded parallel flows, turbulent transition takes place at a critical value of the energy gradient parameter, $K_{max}$, about 370-380, below which no turbulence exists. (3) The location where the flow instability is first initiated accords with the experiments. This location is at y/h=0.58 for plane Poiseuille flow and at r/R=0.58 for pipe Poiseuille flow, which have been confirmed by Nishioka et al (1975)'s experiments and Nishi et al(2008)'s experiments, respectively.

The physical implication of the critical Reynolds number for turbulence transition can then be reinterpreted from this result. Since the flow instability and the initial transition to turbulence can be described by the "Energy Gradient Method," it is reasonable to deduce that the coherent structure in developed turbulence is dominated by the variation of energy gradient and energy loss of mean flows. Furthermore, the turbulence could be controlled by manipulating the energy gradient and energy loss.

Although the analytical model may be simply a heuristic one, the results found in present research are very inspiring. Further work following this streamline may reveal significant findings in this field.

**Acknowledgement**

The author is grateful to Prof. LN. Trefethen (Oxford University) for helpful discussions. He also wants to thank Dr. David Whyte (IHPC in Singapore) for his help in the revision of this paper.

# References

Adrian, R.J., Meinhart, C.D., Tomkins, C.D.,2000, Vortex organization in the outer region of the turbulent boundary layer, J. Fluid Mech., 422, 1-54.

Chapman, S.J., 2002, Subcritical transition in channel flows, J. Fluid Mech., 451, 35-97.

Darbyshire, A.G., and Mullin, T., 1995, Transition to turbulence in constant-mass-flux pipe flow, J. Fluid Mech., 289, 83-114.

Daviaud, F., Hegseth,J., and Berge, P., 1992, Subcritical transition to turbulence in plane Couette flow, Phys. Rev. Lett., 69, 2511-2514.

Drazin, P.G., and Reid,W.H.,2004, Hydrodynamic Stability, Cambridge University Press, Cambridge, 2nd Edition.

Dou, H-S., 2004, Viscous instability of inflectional velocity profile, Proceedings of the Forth International Conference on Fluid Mechanics, Ed. by F. Zhuang and J. Li, Tsinghua University Press & Springer-Verlag, July 20-23, Dalian, China, pp.76-79.

Dou, H.-S., 2006, Mechanism of flow instability and transition to turbulence, International Journal of Non-Linear Mechanics, 41, 512-517.

Dou, H.-S., 2007, Three important theorems for flow stability, Proceedings of the Fifth International Conference on Fluid Mechanics, Ed. by F. Zhuang and J. Li, Tsinghua University Press & Springer, pp. 57-60.

Dou, H-S., Khoo, B. C., 2009, Mechanism of wall turbulence in boundary layer flows, Modern Physics Letters B, 23 (3), 457-460.

Dou, H.-S., Khoo, B.C., 2010, Criteria of turbulent transition in parallel flows, Modern Physics Letters B, 24 (13), 1437-1440.

Dou, H.-S., Khoo, B.C., 2011, Investigation of Turbulent transition in plane Couette flows Using Energy Gradient Method, Advances in Appl. Math. and Mech., 3(2), 165-180. http://arxiv.org/abs/nlin.CD/0501048

Dou, H.-S., Khoo, B.C., and Yeo, K.S., 2007, Energy Loss Distribution in the Plane Couette Flow and the Taylor-Couette Flow between Concentric Rotating Cylinders, International Journal of Thermal Science, 46, 262-275. http://arxiv.org/abs/physics/0501151

Dou, H.-S., Khoo, B.C., and Yeo, K.S.,2008, Instability of Taylor-Couette Flow between Concentric Rotating Cylinders, International Journal of Thermal Science, 47, 1422-1435. http://arxiv.org/abs/physics/0502069

Einstein, A., 1956, Investigations on the Theory of Brownian Movement. New York: Dover.

Emmons, H.W., 1951, The laminar-turbulent transition in a boundary layer -Part 1, J. of Aero. Sciences, 18, 490-498.

Govindarajan, R., Narasimha, R., 1991,The role of residual nonturbulent disturbances on transition onset in 2-dimensional boundary-layers, ASME Journal of fluids engineering, 113, 147-149.

Grossmann, S., 2000,The onset of shear flow turbulence. Reviews of Modern Physics, 72, 603-618.

Hof, B., Juel, A., and Mullin, T., 2003, Scaling of the turbulence transition threshold in a pipe, Physical Review letters, 91, 244502.

Hof,B., van Doorne,C.W.H., Westerweel, J., Nieuwstadt, F.T.M., Faisst, H.,
Eckhardt, B., Wedin,H., Kerswell,R.R., Waleffe,F., 2004, Experimental observation of nonlinear traveling waves in turbulent pipe flow, Science, 305, No.5690, 1594-1598.

Hommema, S.E., and Adrian, R.J., 2002, Similarity of apparently random structure in the outer region of wall turbulence, Experiments in Fluids, 33, 5-12.

Jimenez, J., and Pinelli, A., 1999, The autonomous cycle of near-wall turbulence, J. Fluid Mech., 389, 335-359.

Kline, S.J., Reynolds, W.C., Schraub, F.A., and Runstadler, P.W., 1967, The structure of turbulent boundary layers, J. Fluid Mech. 30, 741-773.

Lee, C.B., 2000, Possible universal transitional scenario in a flat plate boundary layer: Measurement and visualization, Physical Review E, 62, 3659-3670.

Lessen, M., and Cheifetz, M.G.,1975, Stability of plane Coutte flow with respect to finite two-dimensional disturbances, Phys. Fluids, 18, 939-944.

Lin, C.-C., 1955, The Theory of Hydrodynamic Stability, Cambridge Press, Cambridge, 1-153.

Malerud, S., Mälfy, K. J., and Goldburg, W. I., 1995, Measurements of turbulent velocity fluctuations in a planar Couette cell, Phys. Fluids, 7, 1949-1955.

Nishi,M.,Unsal,B., Durst, F., and Biswas, G., 2008, Laminar-to-turbulent transition of pipe flows through puffs and slugs, J. Fluid Mech., 614, 425̇C446.

Nishioka, M., Iida, S., and Ichikawa, Y., 1975, An experimental investigation of the stability of plane Poiseuille flow, J. Fluid Mech., 72, 731-751.

Patel, V.C., and Head, M.R., 1969, Some observations on skin friction and velocity profiles in full developed pipe and channel flows, J. Fluid Mech., 38, 181-201.

Peixinho, J., and Mullin, T., 2007, Finite-amplitude thresholds for transition in pipe flow, J. Fluid Mech. 582, 169-178.

Perry, A.E., and Chong, M.S., 1982, On the mechanism of wall turbulence, J. Fluid Mech. 119, 173-217.

Pfenninger, W.: Boundary layer suction experiments with laminar flow at high Reynolds numbers in the inlet length of a tube by various suction methods. In: G.V. Lachmann (ed.) Boundary layer and flow control, vol. 2, pp. 961–980. Pergamon (1961)

Reynolds, O., 1883, An experimental investigation of the circumstances which determine whether the motion of water shall be direct or sinuous, and of the law of resistance in parallel channels, Phil. Trans. Roy. Soc. London A, 174, 935-982.

Robinson, S.K., 1991, Coherent motion in the turbulent boundary layer. Annu. Rev. Fluid Mech., 23, 601-639.

Shan, H., Zhang, Z., Nieuwstadt, F.T.M., 1998, Direct numerical simulation of transition in pipe flow under the influence of wall disturbances, Inter. J. of Heat and Fluid Flow, 19, 320-325.

Singer,B.A., Joslin, R.D.,1994, Metamorphosis of a hairpin vortex into a young turbulent spot, Physics of Fluids, 6, 3724-3736.

Theodorsen, T., 1952, Mechanism of Turbulence, Proc. 2nd Midwestern Conf. on Fluid Mechanics, Ohio State University, Columbus, OH,1-18.

Tillmark, N., and Alfredsson, P.H., 1992, Experiments on transition in plane Couette flow, J. Fluid Mech., 235, 89-102.

Trefethen, L.N., Trefethen, A.E., Reddy, S.C., and Driscoll, T.A., 1993, Hydrodynamic stability without eigenvalues, Science, 261, 578-584.

Waleffe, F.,1995, Transition in shear flows, nonlinear normality versus nonnormal linearity, Phys. Fluids, 7, 3060-3066.

Waleffe, F., 1997, On a self-sustaining process in shear flows, Phys. Fluids, 9, 883-900.

White, F.M., 1991, Viscous fluid flow, 2nd Edition, McGraw-Hill, New York, 335-337.

Wu, J .-Z., Ma, H.-Y., Zhou, M.-D., 2006, Vorticity and Vortex Dynamics, Springer, Berlin.

| Flow type | Re expression | Eigenvalue analysis, $Re_c$ | Energy method $Re_c$ | Experiments, $Re_c$ | Energy gradient Method, $K_{max}$ at $Re_c$ (from experiments), $\equiv K_c$ |
|---|---|---|---|---|---|
| Pipe Poiseuille | $Re = \rho UD/\mu$ | Stable for all Re | 81.5 | 2000 | 385 |
| Plane Poiseuille | $Re = \rho UL/\mu$ | 7696 | 68.7 | 1350 | 389 |
|  | $Re = \rho u_0 h/\mu$ | 5772 | 49.6 | 1012 | 389 |
| Plane Couette | $Re = \rho Uh/\mu$ | Stable for all Re | 20.7 | 370 | 370 |

Table 1 Comparison of the critical Reynolds number and the energy gradient parameter $K_{max}$ for plane Poiseuille flow and pipe Poiseuille flow as well as for plane Couette flow (Dou,2006). *U* is the averaged velocity, $u_0$ the velocity at the mid-plane of the channel, *D* the diameter of the pipe, *h* the half-width of the channel for plane Poiseuille flow (L=2h) and plane Couette flow. The experimental data for plane Poiseuille flow and pipe Poiseuille flow are taken from Patel and Head (1969). The experimental data for plane Couette flow is taken from Tillmark and Alfredsson (1992), Daviaud et al (1992), and Malerud et al (1995). Here, two Reynolds numbers are used since both definitions are employed in literature. The data of critical Reynolds number from energy method are taken from Drazin and Reid (2004).

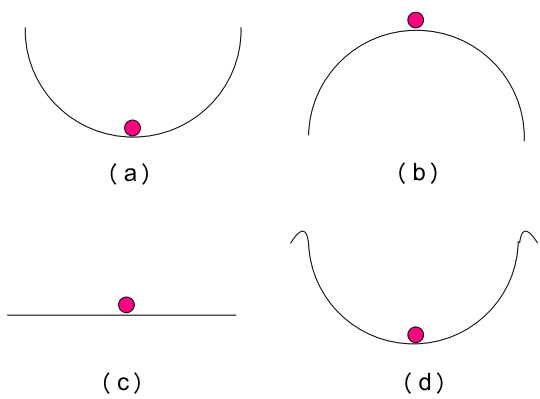

Fig.1 Schematic of equilibrium states of a mechanical system. (a) Stable; (b)Unstable; (c) Neutral stable; (d) Nonlinear unstable (stable for small disturbance but unstable for large ones).

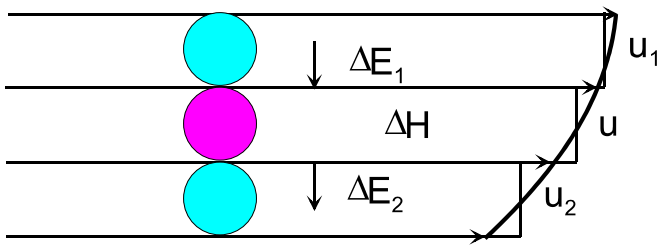

Fig.2 Schematic of fluid particle flows in a steady parallel shear flow.

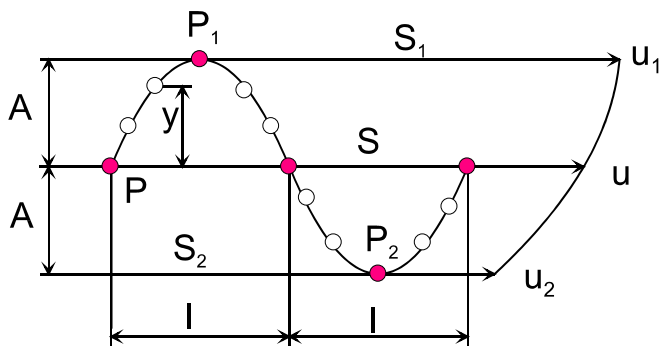

Fig.3 Movement of a particle around its original equilibrium position in a cycle of disturbance.

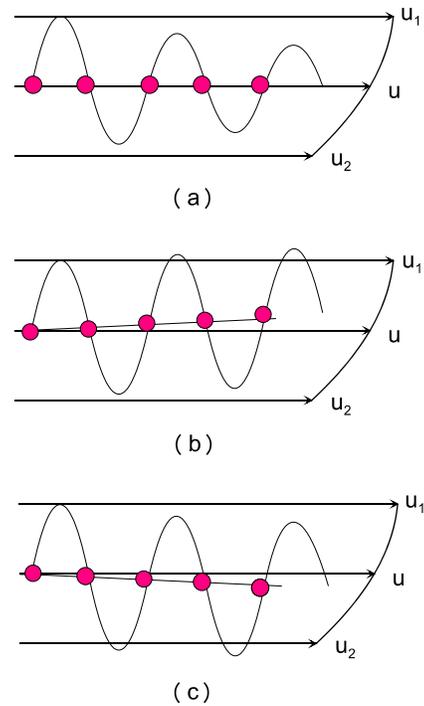

Fig.4 Description of stability of a particle using the energy argument. (a) Stable owing to the energy variation not to exceed the threshold; (b) Losing its stability by gaining more energy; (c) Losing stability by releasing more energy.

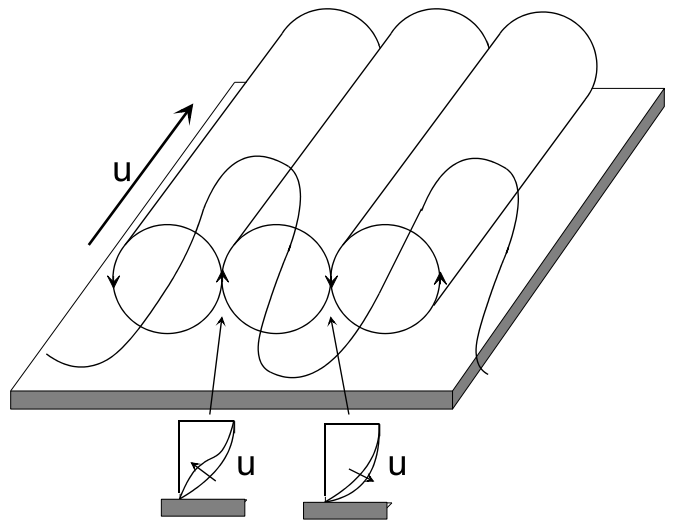

Fig.5 Streamwise vortices makes the velocity profile periodically inflectional and swollen and formation of hairpin vortices.

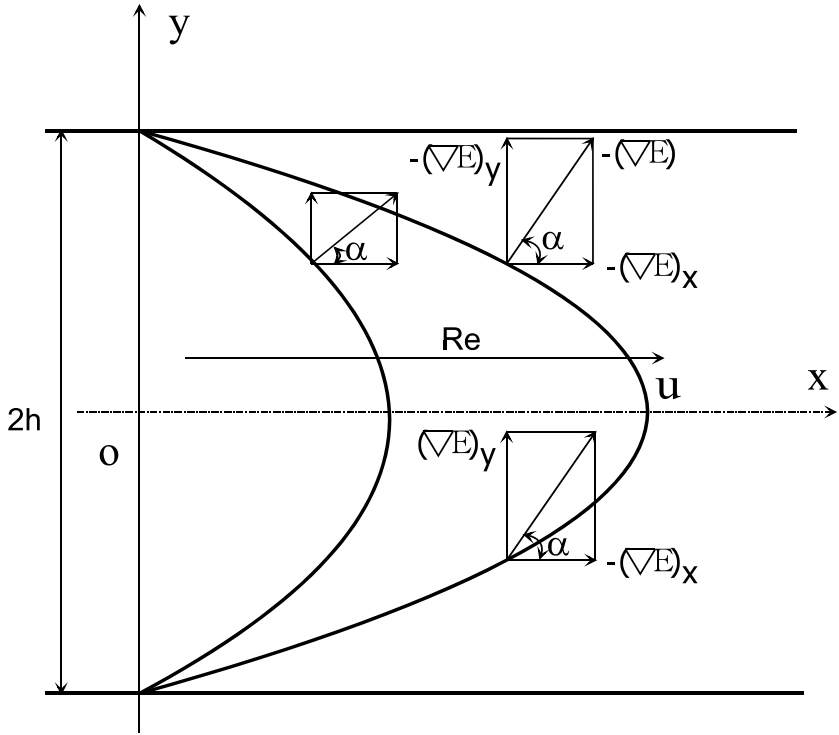

Fig.6 Schematic of energy gradient and energy angle for plane Poiseuille flows.

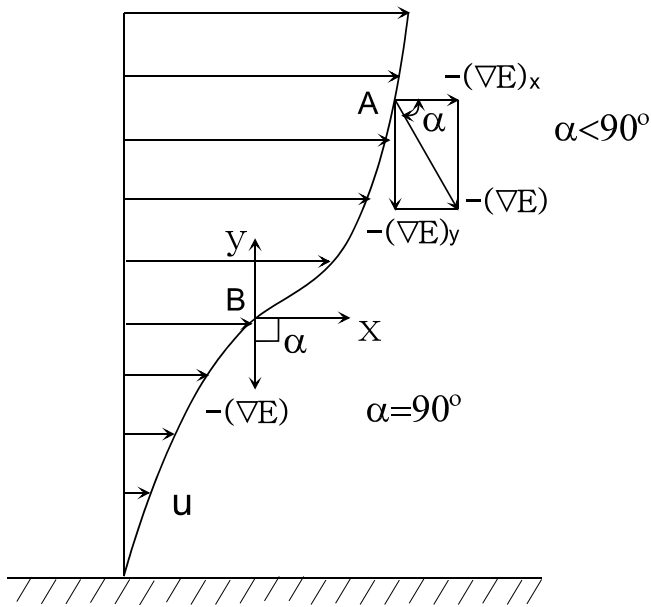

Fig.7 Schematic of the direction of the total energy gradient and energy angle for flow with an inflection point at which the energy angle equals 90 degree.

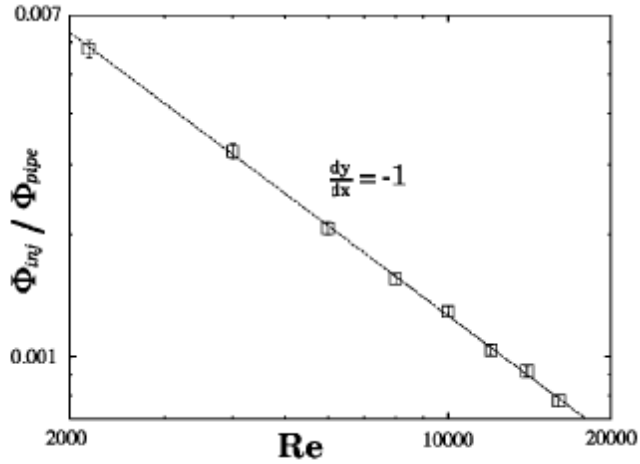

Fig.8 Experimental results for pipe flow: the normalized flow rate of disturbance versus the Reynolds number (Hof, Juel, and Mullin (2003)). The range of Re is from 2000 to 18,000. The normalized flow rate of disturbance is equivalent to the normalized amplitude of disturbance for the scaling of Reynolds number, $\Phi_{inj}/\Phi_{pipe} \sim (v'_m/U)_c$.

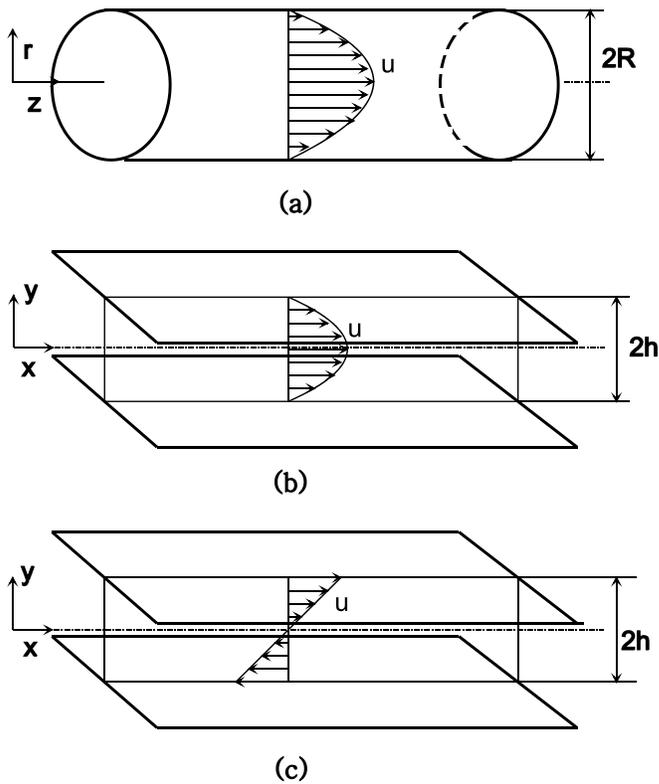

Fig.9 Schematic of wall bounded parallel flows. (a) Pipe Poiseuille flow; (b) Plane Poiseuille flow; (c) Plane Couette flow.

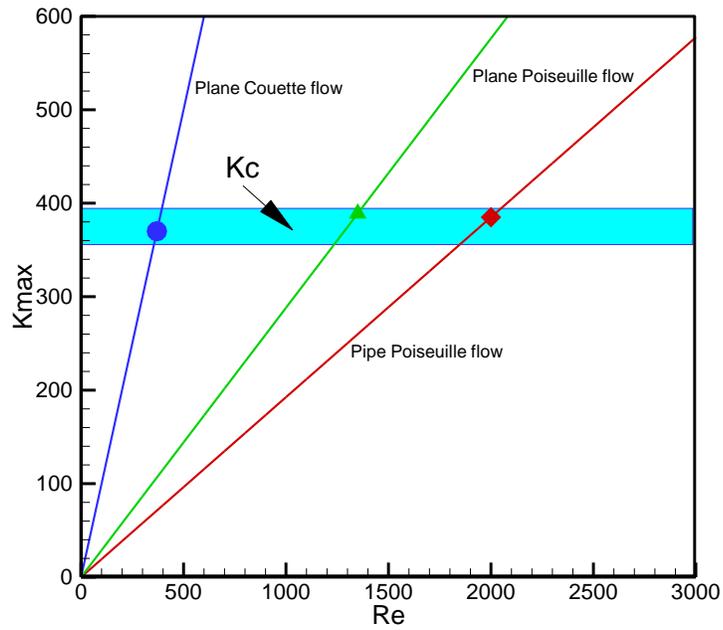

Fig.10 Parameter $K_{max}$ at turbulent transition versus the Reynolds number for various flows. The symbols in the figure represent the data determined from experimental data. For Plane Poiseuille flow, the Reynolds number $\mathrm{Re} \equiv \dfrac{\rho U L}{\mu}$ is used in this figure. The definition of Re is shown in Table 1. The critical value of $K_{max}$ is same for all the wall bounded parallel flows and regardless of the Reynolds number.

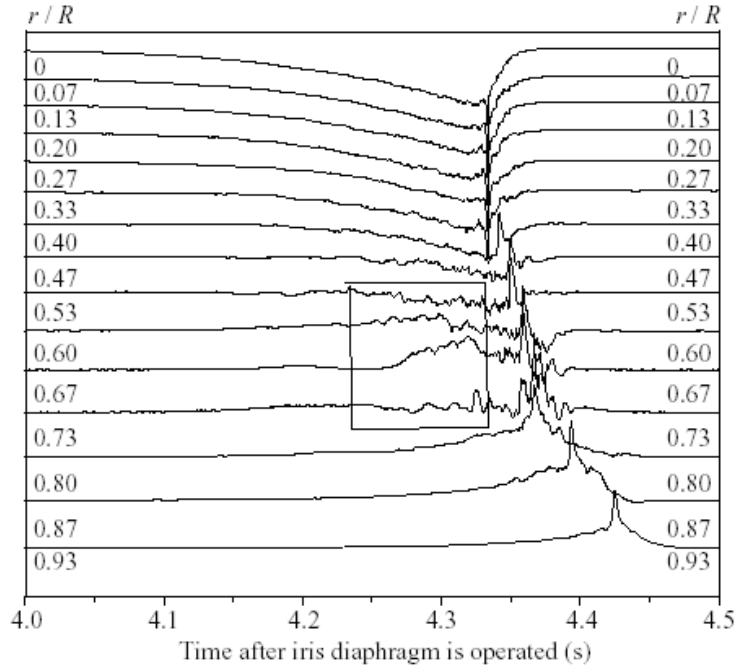

Fig. 11 Axial velocity at different radial position r/R vs. time which is shown from the time when the iris diaphragm is operated at Re=2450, reproduced from [35] (Courtesy of F. Durst; Use permission by Cambridge University Press). The oscillation first started in r/R=0.53~0.73.

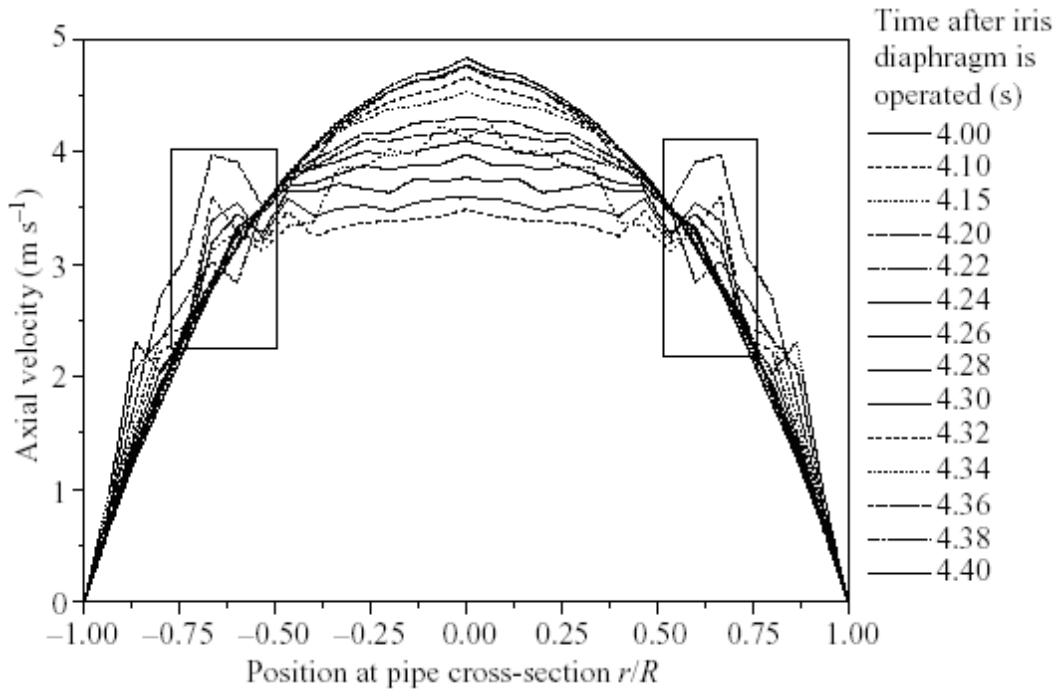

Fig. 12 Axial velocity as a function of different radial position r/R at different time after the iris diaphragm is operated at Re=2450, reproduced from [35] (Courtesy of F. Durst; Use permission by Cambridge University Press). The oscillation first started in r/R=0.53~0.73.